\documentclass[twocolumn,showpacs,showkeys,preprintnumbers,amsmath,amssymb,pra]{revtex4}

\usepackage{graphicx}
\usepackage{dcolumn}
\usepackage{bm}

\newcommand{\bv}[1]{\mbox{\boldmath$#1$}}

\begin{document}


\preprint{ }

\title{Topological Vortex Formation in BEC under Gravitational Field}

\author{Yuki Kawaguchi, Mikio Nakahara$^{1}$,
and Tetsuo Ohmi\\}

\affiliation{%
Department of Physics, Kyoto University, Kyoto 606-8502, Japan\\
$^1$Department of Physics, Kinki University, Higashi-Osaka 577-8502, Japan\\
}%

\date{\today}

\begin{abstract}
Topological phase imprinting is a unique technique for vortex 
formation in a Bose-Einstein condensate (BEC) of alkali metal gas,
in that it does not involve rotation: BEC is trapped in a quadrupole
field with a uniform bias field which is reversed adiabatically leading
to vortex formation at the center of the magnetic trap.
The scenario has been experimentally
verified by MIT group employing $^{23}$Na atoms.
Recently similar experiments have been conducted at Kyoto University, in which
BEC of $^{87}$Rb atoms has been used. In the latter experiments they
found that the fine-tuning of the field reverse time $T_{\rm rev}$
is required to
achieve stable vortex formation. Otherwise, they often
observed vortex fragmentations or a condensate without a vortex.
It is shown in this paper that this behavior
is attributed to the heavy mass of the Rb atom. 
The confining potential, which depends
on the eigenvalue $m_B$ of the hyperfine spin $\bv{F}$ along the magnetic
field, is now shifted by the gravitational field perpendicular to the
vortex line. 
Then the positions of two weak-field-seeking states 
with $m_B=1$ and 2 deviate from each other. This effect is more prominent
for BEC with a heavy atomic mass, for which the deviation is greater and,
moreover, the Thomas-Fermi radius is smaller. 
We found, by solving the Gross-Pitaevskii equation numerically, that
two condensates interact in a very complicated way leading to fragmentation
of vortices, unless $T_{\rm rev}$ is properly tuned.

\end{abstract}

\maketitle

\section{Introduction}

Alkali atom gas undergoes phase transition to a superfluid phase
at very low temperatures ($\sim 0.1~\mu$K) to form a
Bose-Einstein condensate (BEC) \cite{rf:1, rf:2, rf:3, rf:4}. This new BEC is
different from the conventional superfluid $^4$He in that
(i) it is a weakly coupled system for which the Gross-Pitaevskii equation
is applicable with high precision at low temperatures (ii)
alkali atoms
have hyperfine spin degrees of freedom ${\bv{F}}$ and, accordingly,
the order parameter has $2F+1$ components \cite{rf:5, rf:6} while 
$^4$He atoms have no such internal structure
(iii) external magnetic field couples with the hyperfine spin,
allowing for easy control of the condensate order parameter,
among others.
For example, the atoms $^{23}$Na and $^{87}$Rb have internal states $F=1$ 
and $F=2$, for which the number of components is
3 and 5, respectively. 

Associated with BEC is a quantized vortex, which demonstrates
the existence of coherent property and superfluidity of the condensate.
The study of vortices in BEC of alkali atoms started as soon as the
BEC was discovered and several scenarios for vortex formation were
proposed, namely (i) dynamical phase imprinting \cite{rf:xa1, rf:xa2}, (ii) optical spoon \cite{rf:xb1, rf:xb2, rf:xb3}
(iii) rotating asymmetric trap \cite{rf:xc1, rf:xc2}
and (iv) topological phase imprinting \cite{rf:7, rf:8, rf:9, rf:10, rf:11}.
We will be concerned in the present paper with a vortex formation in
the last scenario under the gravitational field. 

In topological phase imprinting, the BEC is confined in a quadrupole
magnetic field and a bias field orthogonal to it.
As the bias field is adiabatically reversed, the condensate spin follows
the direction of the local magnetic field and position dependent order
parameter phase is imprinted in BEC in
such a way that a vortex is created along the axis of the trap
when the bias field is reversed. This scenario has been
verified in beautiful experiments at MIT and reported
in Leanhard {\it et. al.} \cite{rf:11},
in which $F=1$ and $F=2$ hyperfine states of $^{23}$Na have been utilized.
The analysis based on Berry's phase \cite{rf:12, rf:9}, 
as well as numerical solutions \cite{rf:8, rf:9, rf:10},
predicts that a vortex thus created has the winding number $2F$, which was 
also verified in \cite{rf:11}.

In recent experiments conducted at Kyoto University \cite{rf:13}, where BEC of
$^{87}$Rb has been used, they observed that fine-tuning of the field reverse
time $T_{\rm rev}$ is required for topological formation of 
a stable vortex. Vortex fragmentation or no vortex has been observed for
different choices of $T_{\rm rev}$. 
We show, in this paper, the existence of a narrow window in $T_{\rm rev}$ for
successful formation of a stable vortex
is attributed to a subtle role played by the gravitational field and
the heavier mass of $^{87}$Rb compared to $^{23}$Na.

This paper is organized as follows. In Sec. II, we outline the
formalism of the problem based on the multi-component spinor 
Gross-Pitaevskii equation. Section III is devoted to the
numerical results which verify our statements.
It is shown that our $T_{\rm rev}$ for stable vortex formation
approximately reproduces that obtained in Kyoto experiments.
Conclusions and discussion are given in Sec. IV.

\section{Topological Phase Imprinting Under Gravitational Field}

We will be mostly concerned with BEC of $^{87}$Rb in $F=2$ hyperfine state.
The case $F=1$ may be analyzed in a similar manner with considerably
less effort. We first introduce the order parameter for $F=2$ BEC and
the Gross-Pitaevskii equation which describes the dynamics of the
condensate \cite{rf:14, rf:15}.
Topological vortex formation in the presence of a gravitational field
is then discussed in detail.

\subsection{Order Parameter}

Let us consider a uniform BEC with hyperspin $F=2$. 
The internal state of the condensate
may be classified according to the eigenvalue $m_z$ of the operator $F_z$,
where $-2 \leq m_z \leq 2$. The corresponding eigenvector is denoted
as $|m_z \rangle$: $F_z|m_z \rangle = m_z |m_z \rangle$.
As for the eigenvectors we employ the following convention
\begin{equation}
\begin{array}{l}
|2 \rangle= (1,0,0,0,0)^T,\ |1 \rangle = (0,1,0,0,0)^T,\vspace{1mm}\\
|0 \rangle
= (0,0,1,0,0)^T,\ |-1 \rangle = (0,0,0,1,0)^T,\vspace{1mm}\\
|-2 \rangle =(0,0,0,0,1)^T,
\end{array}
\label{eq:base}
\end{equation}
where $T$ stands for transpose operation. An arbitrary order parameter
is then expanded as
\begin{equation}
|\Psi \rangle = \sum_{m_z=-2}^2 \Psi_{m_z}|m_z \rangle = 
(\Psi_2, \Psi_1, \Psi_0, \Psi_{-1}, \Psi_{-2})^T.
\end{equation}

In case the external magnetic field $\bv{B}$ is position-dependent,
the classification of the hyperfine state in terms of $m_z$ is not
appropriate and, instead, the eigenstates of 
$F_{B}\equiv \bv{F}\cdot \bv{B}/B$ must be utilized. 

Consider the order parameter
\begin{equation}
|\Psi \rangle_0 = f_0 |2 \rangle,
\label{eq:b0}
\end{equation}
which satisfies the eigenvalue equation $F_z|\Psi \rangle_0 
= 2 |\Psi \rangle_0$. The corresponding eigenstate with
respect to an arbitrary $F_B$, $F_B|\Psi \rangle = 2 |\Psi \rangle$,
with
\begin{equation}
\bv{B}(\mathbf{r})= B \left( \begin{array}{c}
\sin \beta \cos \alpha\\
\sin \beta \sin \alpha\\
\cos \beta
\end{array} \right)
\label{eq:mag}
\end{equation}
is obtained by rotating $|\Psi_0 \rangle$ through
Euler angles $\alpha, \beta$ and $\gamma$. Explicit calculation shows
that 
\begin{eqnarray}
|\Psi \rangle&=&
\exp(-i \alpha F_z) \exp(-i \beta F_y) \exp(-i\gamma F_z) |\Psi_0 \rangle
\nonumber\\
&= &f_0 e^{-2 i\gamma} \left( \begin{array}{c}
e^{-2i \alpha} \cos^4 \frac{\beta}{2}\vspace{1mm}\\
2 e^{-i \alpha} \cos^3 \frac{\beta}{2}\sin \frac{\beta}{2}\vspace{1mm}\\
\sqrt{6} \cos^2 \frac{\beta}{2}\sin^2 \frac{\beta}{2}\vspace{1mm}\\
2 e^{i \alpha} \cos \frac{\beta}{2}\sin^3 \frac{\beta}{2}\vspace{1mm}\\
e^{2i \alpha} \sin^4 \frac{\beta}{2}
\end{array}
\right) \equiv f_0 |2\rangle_B,
\label{eq:wfss}
\end{eqnarray}
where the last equality defines $|2 \rangle_B$.
This state is under an attractive force toward a region with 
smaller $B=|\bv{B}|$ and hence is called a weak-field-seeking 
state (WFSS). We also obtain $| 1 \rangle_B$ starting from
$|1 \rangle$ in Eq. (\ref{eq:base}), which is also called a
weak-field-seeking state. The states $| -1 \rangle_B$
and $|-2 \rangle_B$, obtained from
$|-1 \rangle$ and $|-2 \rangle$, respectively, are called
strong-field-seeking states (SFSS) while the state $|0 \rangle_B$
is independent of the magnetic field
and will be called the neutral state (NS).

\subsection{Gross-Pitaevskii Equation}

The dynamics of $F=2$ BEC in the limit of zero temperature is given,
within the mean field approximation, by the time-dependent
Gross-Pitaevskii (GP) equation which is extended so that
the spin degrees of freedom are taken into account as \cite{rf:14}
\begin{eqnarray}
   i\hbar\frac{\partial}{\partial t}\Psi_{m} &=& \left[-\frac{\hbar^2}{2M}
   \nabla^2 -M g x \right] \Psi_m \nonumber\\
& & +g_1|\Psi_n|^2 \Psi_m 
   + g_2\left[\Psi_{n}^\dagger (F_{k})_{np} \Psi_{p}\right]
   (F_{k})_{mq}\Psi_q \nonumber\\
   & &+ 5g_3\Psi_n^\dagger \langle 2m2n|00\rangle
   \langle 00|2p2q\rangle \Psi_p\Psi_q \nonumber\\
& &   +\frac{1}{2}\hbar\omega_{Lk}(F_k)_{mn}\Psi_n,
\label{eq:gptime}
\end{eqnarray}
where summations are over $k=x,y,z$ and $-2 \leq n,p,q \leq 2$.
Here, $M$ is the mass of the atom and the gravitational force is assumed
to be along the positive $x$ axis.
The Larmor frequency is defined as $\hbar\omega_{Lk}=\gamma_\mu B_k$, where
$\gamma_\mu \simeq \mu_B$ is the gyromagnetic ratio of the atom,
$\mu_B$ being the Bohr magneton. The interaction parameters $\{g_i\}$ are
expressed in terms of the $s$-wave
scattering length $a_F$, $F$ being the total
hyperfine spin of the two-body scattering state, and are given by
\cite{rf:14}
\begin{eqnarray}
g_1 &=& \frac{4\pi\hbar^2}{M}\frac{4a_2+3a_4}{7} \nonumber \\
g_2 &=& -\frac{4\pi\hbar^2}{M}\frac{a_2-a_4}{7} \nonumber \\
g_3 &=&
 \frac{4\pi\hbar^2}{M}\left(\frac{a_0-a_4}{5}-\frac{2a_2-2a_4}{7}\right).
\end{eqnarray}
The parameters are
$a_0=4.73$\;nm, $a_2=5.00$\;nm and $a_4=5.61$\;nm for $^{87}$Rb atoms.

\subsection{Topological Phase Imprinting}

Suppose a WFSS with $m_B = 2$ is placed in a Ioffe-Pritchard trap.
We consider a condensate uniform along the $z$ axis and confined 
within the $xy$ plane by the quadrupole magnetic field
$$
\bv{B}_{\perp}(\bv{r}) = \left(
\begin{array}{c}
B_{\perp}(r) \cos(-\phi)\\
B_{\perp}(r) \sin (-\phi)\\
0
\end{array} \right)
$$
where $B_{\perp}(r)$ is, to a good approximation,
linear in $r$ within the Thomas-Fermi radius; $B_{\perp}(r) = B' r$.
Here $(r, \phi, z)$ are the cylindrical coordinates where the axis of
the magnetic trap is taken as the $z$ axis.
To prevent the condensate from escaping from the trap through Majorana
flips, a uniform bias field $\bv{B}_z(t)$ must be applied along the $z$ axis,
where we have explicitly written the $t$ dependence of the bias field, which
is necessary for topological phase imprinting.
The total magnetic field is thus given by
\begin{equation}
\bv{B}(\bv{r}, t) = \bv{B}_{\perp}(\bv{r})+\bv{B}_z(t)=\left(
\begin{array}{c}
B_{\perp}(r) \cos(-\phi)\\
B_{\perp}(r) \sin (-\phi)\\
B_z(t)
\end{array} \right).
\end{equation}
The condensate would be cylindrically symmetric 
around the $z$ axis if it were not for the gravitation field.
In the presence of the 
gravitational field with a potential $-Mgx$, in contrast, 
the condensate is not symmetric any more. 

The initial condensate profile is determined by solving the stationary
GP equation, in which the $m_B=2$ WFSS (\ref{eq:wfss}) is substituted.
The angles in (\ref{eq:wfss}) are now given by
\begin{equation}
\begin{array}{l}
\displaystyle \alpha = -\phi=-\tan^{-1} \frac{y}{x},\vspace{.2cm}\\
\displaystyle \beta = 
\tan^{-1}\frac{B_{\perp}(r)}{B_z(t)}=\tan^{-1}\frac{B'\sqrt{x^2+y^2}}{B_z(t)}.
\end{array}
\end{equation}
It is assumed that the initial state is vortex free, which forces us to 
choose $\alpha = -\gamma (=-\phi)$ in Eq. (\ref{eq:wfss}). 
The ground state condensate profile corresponds to the lowest energy
eigenvalue of the equation
\begin{eqnarray}
\left[-\frac{\hbar^2}{2M}\nabla^2-Mgx\right](f_0 v_m)&+&(g_1+ 4 g_2) f_0^3 v_m
\nonumber\\
+ \hbar \omega_L f_0 v_m &=& \mu f_0 v_m,
\end{eqnarray}
where $v_m = \langle m|2 \rangle_B$,
$\nabla^2=\partial^2/\partial x^2+\partial^2/\partial y^2$ and
\begin{equation}
\hbar \omega_{L} = \gamma_{\mu} \sqrt{{B'}^2(x^2+y^2) +B_z(t)^2}.
\end{equation}
The eigenvalue $\mu$
is identified with the chemical potential.
If $\{v_m\}^{\dagger}$ is multiplied from the left and
$\sum_m|v_m|^2 =1$ and similar identities are employed, 
we obtain the reduced GP equation for $f_0$;
\begin{eqnarray}
&-&\frac{\hbar^2}{2M} \left[\nabla^2 f_0 + 
\sum_m v_m^* \left\{(\nabla^2 v_m) f_0+2 \nabla v_m \cdot \nabla f_0\right\}
\right]\nonumber\\
& & -Mgx f_0+ (g_1 + 4 g_2) f_0^3 + \hbar \omega_L f_0 = \mu f_0,
\label{eq:gs}
\end{eqnarray}
The initial condensate profile and the chemical potential $\mu$ are 
obtained by solving (\ref{eq:gs}) numerically.

The uniform bias field is then reversed as
\begin{equation}
B_z(t) = \left\{ 
\begin{array}{cl}
\displaystyle 
B_z(0) \left(1-\frac{2t}{T_{\rm rev}}\right)& 0 \leq t \leq T_{\rm rev}
\vspace{.2cm}\\
-B_z(0)& T_{\rm rev} < t.
\end{array} \right.
\end{equation}
It has been shown, in the previous works \cite{rf:9,rf:10}
without taking gravity into account,
that a vortex formation takes place at the center of the condensate
when the bias field has been reversed. The bias field disappears at 
$t\sim T_{\rm rev}/2$ and the $m_B=2$ condensate transforms into 
$-2\le m_B \le 1$ components through Majorana flips. 
The components with $m_B \leq 0$ are not confined and escape from
the trap. The final condensate is then made
of a mixture of the $m_B =1$ and $m_B=2$ WFSS. 
It is found from Eq.~(\ref{eq:wfss}) with $\beta=\pi$
that the vortex in the $m_B=2$ component
has the winding number four while a similar condensate amplitude for
$m_B=1$ tells us that the vortex in this component has the winding
number three \cite{rf:10}. It is essential, therefore, to 
employ the full spinor GP equation (\ref{eq:gptime}) to analyze this 
system. 

In the presence of the gravitational field, however,
the center of the condensate shifts from the axis of the magnetic trap and,
moreover, the minimum of the trapping potential, including gravity,
depends on the hyperfine state due to the last term in (\ref{eq:gptime}).
To be more concrete, the $m_B$ component of the condensate
in the Ioffe-Pritchard trap is subject to the external potential
\begin{equation}
V(x, y, t)= \frac{1}{2} m_B \gamma_{\mu} \sqrt{{B'}^2 (x^2+y^2) + B_z(t)^2} -M g x.
\label{eq:exactpot}
\end{equation}
Note that the center of the quadrupole field is $(0, 0)$ while
the minimum of the potential is
\begin{equation}
(x_0, y_0) \equiv
\left(\frac{2 M g |B_z(t)|}{B'\sqrt{{B'}^2 m_B^2 \gamma_{\mu}^2 - 4 M^2 g^2}}, 
0 \right)
\label{eq:exact}
\end{equation}
where we have assumed that $B' m_B \gamma_{\mu} > 2 M g$, which always
holds in actual experiments. Otherwise, there is no stable minima in
the $x$-direction and the condensate cannot be confined.
Observe that the deviation $x_0$ has the same time dependence
as $|B_z(t)|$ and that it
is larger for $m_B=1$ than for $m_B=2$, implying that
there is a finite distance between the centers of the $m_B=1$ 
and $m_B=2$ condensates. 
It is important to note also that $x_0$ is proportional to the atomic
mass $M$, provided that $B'm_B \gamma_{\mu} \gg 2 Mg$,
and hence the effect of the gravity on the condensate is more
prominent for atoms with heavy mass.

The potential $V(x, y, t)$ is approximated by a displaced harmonic potential 
when $|B_z| \gg |B_{\perp}|$, taking the form
\begin{eqnarray}
V(x, y, t) &\simeq &
\frac{m_B \gamma_{\mu}{B'}^2}{4|B_z(t)|}(x^2+y^2)-Mgx+\frac{1}{2}m_B \gamma_{\mu}|B_z(t)|
\nonumber\\
&=& \frac{1}{2}M \omega_{\rm HO}^2(t)\left[
\left(x-\frac{g}{ \omega_{\rm HO}^2(t)}\right)^2+y^2\right]\nonumber\\& &
- \frac{M g^2}{2 \omega_{\rm HO}^2(t)}+\frac{1}{2}m_B \gamma_{\mu}|B_z(t)|,
\label{eq:harmapp}
\end{eqnarray}
where
\begin{equation}
\omega_{\rm HO}^2(t) = \frac{m_B \gamma_{\mu}}{2 M |B_z(t)|}{B'}^2.
\label{eq:omega2}
\end{equation}
The displacement $x_0= g/\omega_{\rm HO}^2(t)$ found from (\ref{eq:harmapp})
is in agreement with the result (\ref{eq:exact})
when $B'm_B \gamma_{\mu} \gg 2 Mg$.
The frequency $\omega_{\rm HO} \equiv \omega_{\rm HO}(0)$ serves
as an energy scale of the system.
The gravitational potential in Eq.~(\ref{eq:exactpot})  
leads to two remarkable effects on vortex formation.
First, for $F=2$ BEC, there are two WFSS, one ($m_B=1$)
has three units of vorticities while the other ($m_B=2$) has four units.
Their positions deviate due to the difference in $m_B$, which makes the 
interaction between the condensate even more complicated. 
Secondly, the minimum of the potential shifts from the center of the
quadrupole field, where the topological vortex formation takes place.
A highly quantized vortex with winding number $|n| \geq 2$
is unstable against decay into $|n|$ singly quantized vortices.
The lifetime of a highly quantized vortex is longer if its axis is at
the center of the condensate. If it is off-centered, in contrast, 
it is expected that the vortex moves toward the edge of the condensate and
split into $|n|$ singly quantized vortices. 
Therefore a topologically created vortex under gravity will be
unstable against decay unless some caution is exercised, see 
the next section. If the deviation of the center of the quadrupole field
is large compared to the Thomas-Fermi radius, the ``vortex formation''
takes place outside the condensate and it is impossible to observe the
vortex.

These effects are more prominent
in $^{87}$Rb compared to $^{23}$Na. In Kyoto experiments, the parameter
$\omega_{\rm HO} \simeq 2 \pi \times 330\;[{\rm rad/s}]$ is taken to be the 
same as the MIT experiment. Then the Thomas-Fermi radius
$r_{\rm TF}=a_{\rm HO} \sqrt{2\mu/M \omega_{\rm HO}^2}$ has mass dependence
$\sim M^{-1/2}$ through $a_{\rm HO} = \sqrt{\hbar/M \omega_{\rm HO}}$ since
the factor $\mu/M$ roughly takes the same value for both experiments.
%
%

The above consideration shows that the behavior of the condensate depends
heavily on the atomic mass $M$ and the reverse time $T_{\rm rev}$.
We solved the spinor GP equation (\ref{eq:gptime}) numerically for various
$T_{\rm rev}$ for both $^{23}$Na and $^{87}$Rb. In the next section,
we outline mainly the results for the $F=2$ condensate of $^{87}$Rb and then 
compare them with those for $^{23}$Na.

\section{Numerical Results}

We have solved the stationary GP equation (\ref{eq:gs}) numerically
to obtain the vortex-free initial condensate profile $f_0(x, y)$ and the
chemical potential $\mu$. We then solved the time-dependent spinor
GP equation (\ref{eq:gptime}) with $\Psi_m = f_0 v_m$ as the initial
state. 
We had to pay special attention to particles escaping form 
the trap. This is taken care of by introducing an appropriate absorption
potential beyond which the condensate amplitude decays exponentially.
We found slight interference patters due to reflection at the
absorption potential, 
although we believe that this does not cause any qualitative
change in conclusion. 

As for numerical values, we take the following numbers reproducing the Kyoto
experiments: $N = 5 \times 10^5,\ \omega_{\rm HO} = 2\pi \times 330
\;[{\rm Hz}],\ \omega_z= 2\pi \times 17\;[{\rm Hz}]$. The Thomas-Fermi
radius is then $r_{\rm TF} = 3.1\;\mu{\rm m}$. The initial bias field
is $B_z(0)=0.5\;{\rm G}$, which is then reduced to 
$B_z(T_{\rm rev}) = -0.5\;{\rm G}$.

It turns out that a stable vortex formation takes place only 
for $T_{\rm rev}$ in a small window around $T_{\rm rev} \sim 2\;{\rm ms}$.
The behavior of the BEC is qualitatively different for
$T_{\rm rev}$ out of this window from the case $T_{\rm rev} \sim 2\;{\rm ms}$. 
This is in harmony with the observation made by the Kyoto group \cite{rf:13}.
Therefore we take
$T_{\rm rev} = 4\;{\rm ms}, 2\;{\rm ms}$ and $1.5\;{\rm ms}$ as typical values
representing the three classes of $T_{\rm rev}$. 

\subsection{$T_{\rm rev} = 4\;{\rm ms}$}

The change in $B_z$ is slow in this case and each WFSS component
of the condensate follows its equilibrium position (\ref{eq:exact}) as $B_z$
is reversed. There is essentially a single condensate with $m_B=2$
at $t < T_{\rm rev}/2$.
Majorana flips take place at $t \sim T_{\rm rev}/2$ in the vicinity of
the point where the magnetic field vanishes. 
The center of $m_B=2$ condensate is separated from that of $m_B=1$ condensate
from then on.
There exists a vortex with the winding number four (three) in the $m_B=2$ ($m_B=1$) condensate, which are created at the center of the quadrapole field 
at around $t\sim T_{\rm rev}/2$. The center of the condensate is at the
center of the quadrupole field only at $t =T_{\rm rev}/2$ and its
subsequent position follows the minimum of the potential, which depends on
$m_B$. Then these components interact in a complicated manner with each other
and vortex fragmentation takes place.

Figure 1 shows the total density profile $\sum_m|\Psi_m|^2$ 
and the component profile $|\Psi_{m_B}|^2$ for $m_B=1,2$
with $T_{\rm rev}=4\;{\rm ms}$. 


\subsection{$T_{\rm rev} = 2\;{\rm ms}$}

A window for stable vortex formation exists in the vicinity
of $T_{\rm rev} \sim 2\;{\rm ms}$.
The condensate is pushed up by reducing $B_z(t)$
and the center of the condensate reaches the center of the quadrupole field at
$t \simeq T_{\rm rev}/2$ as in the previous case.
The minimum of the potential follows the path it took for $0 < t<
T_{\rm rev}/2$ backward after $t=T_{\rm rev}/2$.
The condensate, however, fails to follow the potential minimum since it has
been accelerated upward. During the acceleration, all atoms are in a single
hyperfine state and equally accelerated. Following the Majorana flips at
$t\simeq T_{\rm rev}/2$,
each component eventually reaches a peak with some delay, after which
it starts to fall chasing for the corresponding potential minimum.
The delay is longer if the acceleration is larger, i.e. $T_{\rm rev}$ is 
smaller, and the strength of the confinement is weaker.
The latter condition depends on $m_B$ as shown explicitly
in Eq.~(\ref{eq:omega2}).

If $T_{\rm rev}$ is fine-tuned to be around 2\;ms,
the distance between the two condensates with $m_B=1$ and $m_B=2$ happens to
be small at $t=T_{\rm rev}$ in spite of separated potential minima for
respective components. Therefore
the $m_B=2$ vortex with the winding number four is not yet disturbed
by the $m_B =1$ vortex with the winding number three, see Fig.~2.
Observe that the center of the $m_B=1$ component
is below (above) that of the $m_B=2$ component at $t=T_{\rm rev}$
in Fig.~1 (Fig.~3).
Vortex fragmentation occurs at $ t > T_{\rm rev}$ 
when the distance between the condensates becomes larger
as in the previous case.

\subsection{$T_{\rm rev} = 1.5\;{\rm ms}$}

Since the change in $B_z(t)$ is fast, the condensate 
cannot follow the minimum of the potential (\ref{eq:exact}) and
does not reach the center of the quadrupole
field at $t=T_{\rm rev}/2$. The center of the vortex
always appears in the center of the quadrupole field and thus
the vortex axis appears above the center of the condensate in this case.
The vortex thus created has the winding number four and a highly
quantized off-centered vortex is expected to decay quickly into four
singly quantized vortices \cite{rf:16}. 
This is in fact manifest in Fig. 3 at $t = T_{\rm rev}$. 
In Kyoto experiments \cite{rf:13},
the wavy edge of the condensate is observed when $T_{\rm rev}$ is smaller
compared to that for stable vortex formation. This might be attributed
to the fragmented vortices.


\section{Conclusions and Discussion}

In summary, we have shown that the effect of the gravitational
field on the topological vortex formation in BEC of heavy atoms,
such as $^{87}$Rb, is not negligible. Unless the reverse time
$T_{\rm rev}$ is fine-tuned, fragmentation of a multiply quantized
vortex takes place during topological phase engineering. This 
fragmentation is attributed to the existence of two WFSSs
with $m_B=1$ and $2$, whose centers are separated, under gravity, by
distance comparable to the Thomas-Fermi radius. WFSS with $m_B=1$
supports a vortex with three units of winding number while that
with $m_B=2$ supports four units.

When the two vortices have the same center,
there exists an axially symmetric stable solution~\cite{rf:10}.
If the vortex centers are displaced from each other,
however, the two WFSSs interfere with each other
through the $g_2$- and $g_3$-terms in Eq.~\eqref{eq:gptime}
so that the deviation from axial symmetry is further amplified,
leading to fragmentation observed in~\cite{rf:13}. 

When the field is reversed so slowly that each hyperfine state follows its potential minimum,
the center of two hyperfine components deviate from each other
as soon as Majorana flips take place at $t\sim T_{\rm rev}/2$,
leading to vortex fragmentation.
When the field is reversed so fast, in contrast,
the condensate fails to follow the potential minimum
and the deviation of two components at $t=T_{\rm rev}$ is minute.
The vortex is, however, imprinted near the edge of the condensate,
which leads to spliting of the multiply quantized vortex into singly quantized vortices.
The fragmentation pattern often has two-fold symmetry,
which originates from the vertical displacement of two condensates~\cite{rf:16}.
To observe a stable multiply quantized vortex,
the field reverse time must be taken somewhere between above two cases.

We have also studied the effect of the gravitational field on the BEC
of $^{23}$Na with $F=2$, which corresponds to the MIT experiments.
We have shown that there exist a similar window of $T_{\rm rev}$
for a topological formation of a stable vortex in this case. 

\section*{Acknowledgements}

We are grateful to Tsutomu Yabuzaki, Mitsutaka Kumakura 
and Takashi Hirotani for informing us
of their experimental results prior to publication.
The computation in this work has been done using
the facilities of the Supercomputer Center, 
Institute for Solid State Physics, University of Tokyo.
YK and TO are supported by the Grant-in-Aid for the 21st Century COE ``Center
for Diversity and Universality in Physic'' from the Ministry of Education,
Culture, Sports, Science and Technology (MEXT) of Japan.
YK would like to acknowledge support from the Fellowship
Program of the Japan Society for Promotion of Science (JSPS) (Project No. 16-0648).
MN is grateful for partial support of a 
Grants-in-Aid for Scientific Research from MEXT (Project No. 13135215) and
JSPS (Project No. 14540346).

\clearpage
\begin{figure*}
\begin{center}
\includegraphics[width=13.5cm]{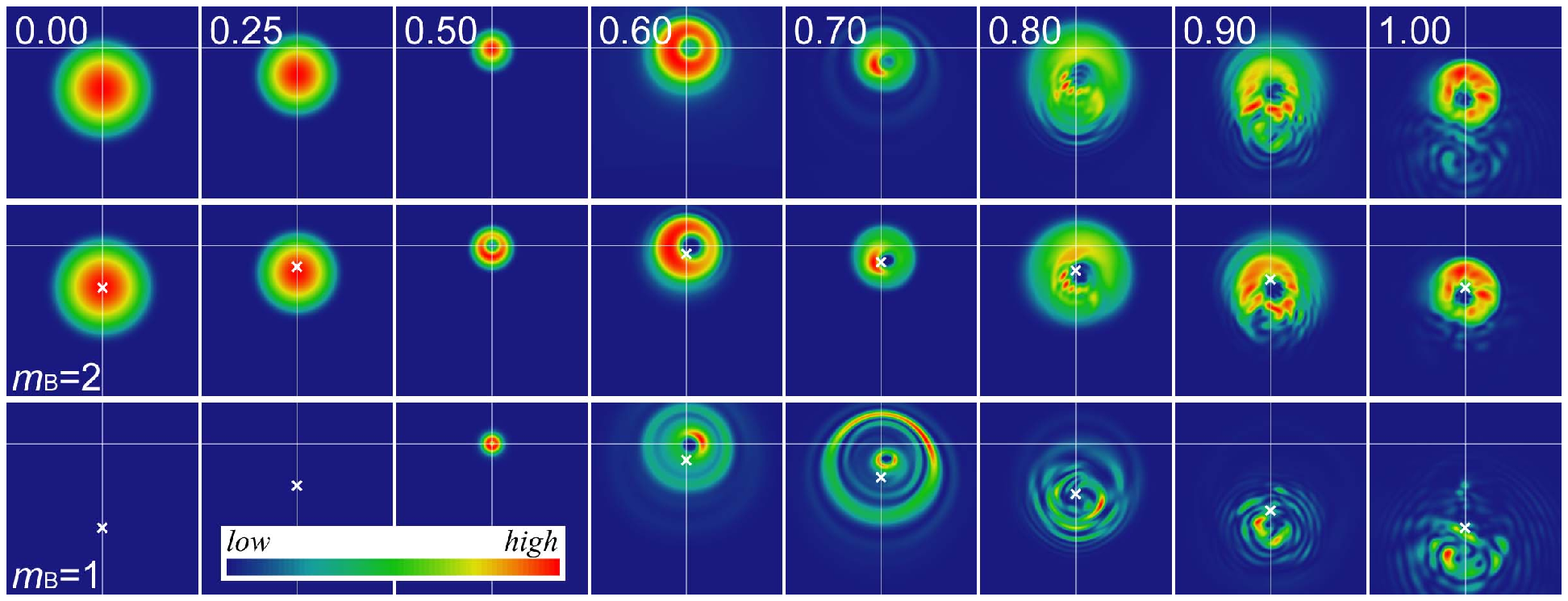}
\vspace{-5mm}
\end{center}
\caption{(Color) Time development of the condensate with $T_{\rm rev}=4$~ms.
The white lines show the coordinate system, the origin of which is
the center of the quadrupole field where the vortex formation takes place.
Upper panel shows that total density of the condensate. The number
in the panel shows the normalized time $t/T_{\rm rev}$.
The middle (lower) panel the density of $m_B =2\ (m_B=1)$ component. 
The cross ($\times$) shows the minimum of 
the potential for the respective component. Observe that the center of
each $m_B$ component follows the minimum of the respective potential and
accordingly the center of the the $m_B=1$ component is below that of
the $m_B=2$ component at $t=T_{\rm rev}$.
}
\label{fig:1}
\end{figure*}
\begin{figure*}
\begin{center}
\includegraphics[width=13.5cm]{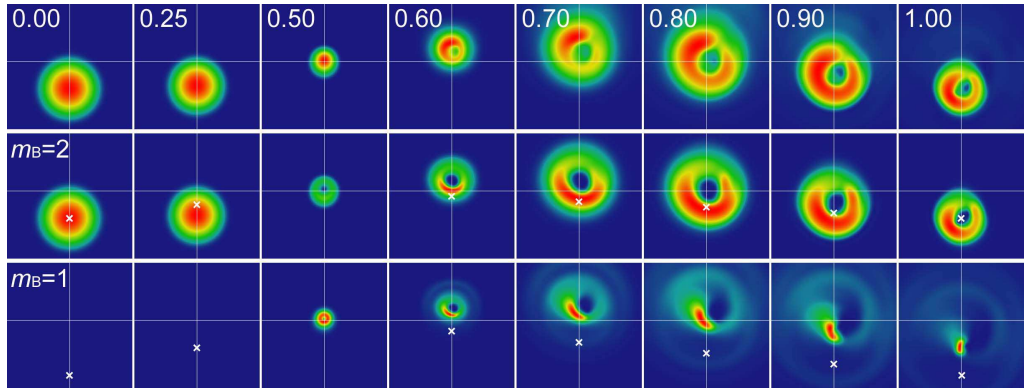}
\end{center}
\vspace{-5mm}
\caption{(Color) Time development of the condensate with $T_{\rm rev}=2$~ms.
See the caption of Fig. 1 for the definition of the symbols.
The condensate is pushed up while the potential is deformed by reversing
$B_z$. The $m_B=1$ component is still above the minimum of its
potential at $t=T_{\rm rev}$ so that the center of the component is close to 
that of the $m_B=2$ component.
}
\label{fig:2}
\end{figure*}
\begin{figure*}
\begin{center}
\includegraphics[width=13.5cm]{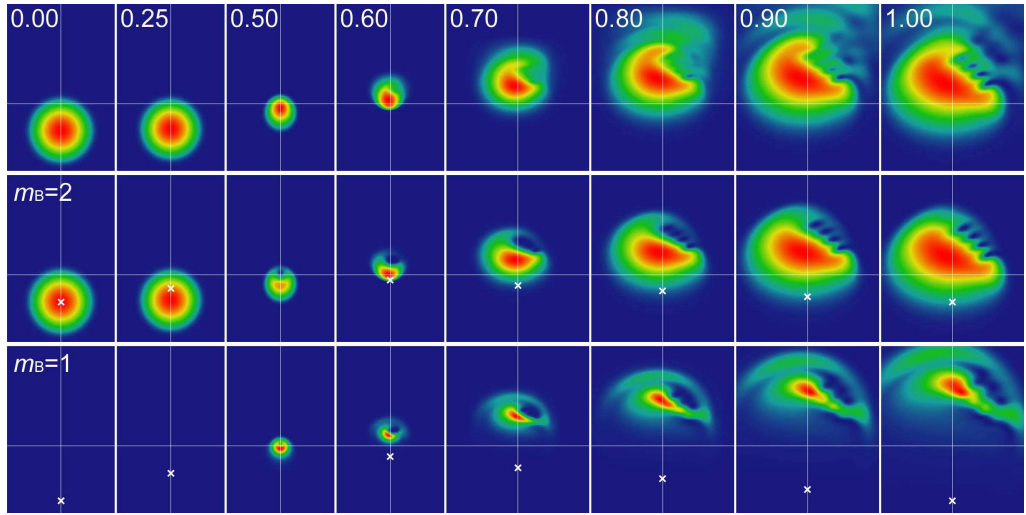}
\end{center}
\vspace{-5mm}
\caption{(Color) Time development of the condensate with $T_{\rm rev}=1.5$~ms.
See the caption of Fig. 1 for the definition of the symbols.
The condensate is pushed up by a quick deformation of the potential 
and the centers of both $m_B$ components are still above the minima
of the potentials at $t=T_{\rm rev}$. This is more eminent for 
$m_B=1$ component so that the center of the $m_B=1$ component is above that of
the $m_B=2$ component.}
\label{fig:3}
\end{figure*}

\end{document}